\documentclass[12pt,preprint]{aastex}

\newcommand{\myemail}{terada@subaru.naoj.org}

\slugcomment{}

\shorttitle{Crystallized Water Ice in Silhouette Disk}
\shortauthors{Terada, H. \& Tokunaga, A. T.}

\begin{document}

\title{Discovery of Crystallized Water Ice in a Silhouette Disk in the M43 Region}

\author{Hiroshi Terada\altaffilmark{1}}
\email{\myemail}

\and

\author{Alan T. Tokunaga\altaffilmark{2}}

\altaffiltext{1}{Subaru Telescope, National Astronomical Observatory
of Japan, 650 North A'ohoku Place, Hilo, HI 96720, USA}

\altaffiltext{2}{Institute for Astronomy, University of Hawaii, 2680
Woodlawn Drive, Honolulu 96822, USA}

\begin{abstract}
We present the 1.9--4.2\,$\micron$ spectra of the five bright (L $\leq$ 11.2) young stars associated with silhouette disks with moderate to high inclination angle of 39--80$\arcdeg$ in the M42 and M43 regions. The water ice absorption is seen toward d121-1925 and d216-0939, while the spectra of d182-316, d183-405, and d218-354 show no water ice feature around 3.1\,$\micron$ within the detection limits. 
By comparing the water ice features toward nearby stars, we find that the water ice absorption toward d121-1925 and d216-0939 most likely originates from the foreground material and the surrounding disk, respectively. 
The angle of the disk inclination is found to be mainly responsible for the difference of the optical depth of the water ice among the five young stars. Our results suggest that there is a critical inclination angle between 65$\arcdeg$ and 75$\arcdeg$ for the circumstellar disk where the water ice absorption becomes strong. The average density at the disk surface of d216-0939 was found to be 6.38 $\times$ 10$^{-18}$ g cm$^{-3}$. The water ice absorption band in the d216-0939 disk is remarkable in that the maximum optical depth of the water ice band is at a longer wavelength than detected before. It indicates that the primary carrier of the feature is purely crystallized water ice at the surface of the d216-0939 disk with characteristic size of $\sim$ 0.8\,$\micron$, which suggests grain growth. This is the first direct detection of purely crystallized water ice in a silhouette disk. 
\end{abstract}

\section{INTRODUCTION}

Planetary systems are thought to evolve from circumstellar disks around young stellar objects. In this context, dynamical and chemical evolution of circumstellar disks has been extensively studied with various observational and theoretical methods to reveal how planets are formed. Recent advanced astronomical facilities of both ground-based 8-10m class telescopes and space missions such as {\it Hubble Space Telescope} ({\it HST}), {\it ISO}, {\it AKARI,} and {\it SPITZER} have revealed morphological and spectroscopic features of circumstellar disks in great detail with high sensitivity and high spatial resolution. 

The most striking discovery was the direct detection of circumstellar disks in Orion Nebula Cluster imaged by $HST$ in its early operation phase \citep{ode93}. They exhibited clear morphology of the circumstellar disks as a silhouette in front of the very bright background nebula illuminated by O, B stars (Trapezium stars) in the cluster. This was a huge milestone for studies of the protoplanetary disks, as it showed that the circumstellar disks could be directly studied with high spatial resolution ($\sim$ 0$\farcs$1). The silhouette disks have definite advantages, because various basic properties of the circumstellar disk such as its shape, size, inclination angle, and composition are unambiguously determined. It provides us with excellent opportunities to investigate the circumstellar disk as a progenitor of the planetary system.

Water is a key element at the top of the hierarchy of building blocks to complex, prebiotic molecules. In our Solar System, water is thought to be brought onto the surface of a terrestrial planet such as the Earth in the form of water ice in its solid phase \citep{boe07}. Therefore, the search for the water ice in circumstellar disks is of extreme interest to investigate the possible emergence of the life elsewhere. The water ice feature at the shortest wavelength appears at 1.5\,$\micron$, 2.0\,$\micron$, and 3.1\,$\micron$ in the near infrared region, which is accessible from the ground. In particular, among those features, the water ice absorption at 3.1\,$\micron$ is the most important in terms of its detectability. 

The water ice absorption at 3.1\,$\micron$ has been detected toward various astronomical sources including young stellar objects since its first detection by \citet{gil73}. While in most of the cases the detected water ice is attributed to the foreground cloud and/or their envelope around the sources, recently the water ice absorption in the circumstellar disks has been detected successfully \citep{pon05,ter07,sch10,aik12}. However, those detections are based on the scattered light by the circumstellar particles around the central star, and the geometry of the water ice absorption is not clear. Water ice absorption seen directly through the silhouette disks will be easier to interpret than those detections. Another advantage comes from the fact that the M42 and M43 regions suffer from less extinction \citep{sca11}, which means that the foreground contamination is minimized for the measurement of the water ice in the circumstellar disk. Taking into account the fact that majority of stars are born in high mass star forming regions, it is important to investigate the closest high mass star forming regions of M42 and M43. The silhouette disks were initially discovered in the Orion Nebular Cluster; M42 \citep{mcc96,bal00}, and recently the sample is largely expanded with several additional silhouette disks found also in M43 \citep{smi05}. In this paper, the distance to M42 is adopted as 414 pc \citep{men07}, while 436 pc is used for M43 \citep{ode08}. 

We present the spectra of five sources with silhouette disks in the M42 and M43 regions, namely d121-1925, d182-332, d183-405, d218-354 \citep{bal00}, and d216-0939 \citep{smi05}. In \S 2, the observational procedures and applied data analysis are described. In \S 3, the continuum of each target is determined and the optical depth of the water ice feature is derived. The origin of the detected water ice, the possible cause of the different optical depth of the water ice, and the interesting nature of the detected water ice in d216-0939 disk are discussed in \S 4. The conclusion is described in \S 5.

\section{OBSERVATION \& DATA REDUCTION}

The combination of near infrared imaging (1.4--4.2\,$\micron$) and spectroscopy (1.9--4.2\,$\micron$) were conducted as a series of the observations from 2000 December to 2007 September.  All the observations were performed under seeing limited conditions using the Infrared Camera and Spectrograph \citep[IRCS;][]{tok98,kob00,ter04} at the Subaru 8.2m Telescope atop Maunakea, Hawaii. IRCS was initially mounted at the Cassegrain focus of the telescope in 2000, and moved onto the Nasmyth focus on 2005 December for the upgraded adaptive optics system \citep[AO188;][]{hay10}. The pixel scales became slightly finer by a factor of 0.89 after moving to the Nasmyth focus. The observation log is shown in Table~\ref{tbl-obslog} with a seeing size on those nights. The seeing values were good in most of the observing runs. But the seeing was poor in some cases, especially for d218-354. Sky conditions throughout all of the observing nights were photometric. 

\begin{deluxetable}{c c c c c c}
	\tabletypesize{\scriptsize}
	\tablecolumns{6}
	\tablewidth{0pc}
    	\tablecaption{Observing Log\label{tbl-obslog}}
	\tablehead{
	\colhead{}&\colhead{Observing}&\colhead{}&\colhead{Exposure}&\colhead{Seeing}&\colhead{Standard}\\
      \colhead{Object}&\colhead{Date}&\colhead{Mode}&\colhead{Time}&\colhead{Size}&\colhead{Star}\\
      \colhead{}&\colhead{(UT)}&\colhead{}&\colhead{(sec)}&\colhead{(\arcsec)}&\colhead{}}
      \startdata
      d121-1925&2001 Mar 02&$L^{\prime}$ Imaging&90&0.58&HD 36512\\
  			&2001 Mar 10&$K$ Spectroscopy&480&0.96&HR 1931\\
      			&2000 Dec 04&$L$ Spectroscopy&2520&0.45&HR1931\\
       d182-332&2001 Mar 02&$L^{\prime}$ Imaging&90&0.66&HD 36512\\
      			&2004 Feb 04&$K$ Spectroscopy&480&0.19&HR 1931\\
      			&2001 Jan 11&$L$ Spectroscopy&1250&0.31&HR 1931\\
       d183-405&2001 Mar 11&$L^{\prime}$ Imaging&765&0.91&HD 36512\\
      			&2004 Feb 04&$K$ Spectroscopy&480&0.32&HR 1931\\
      			&2001 Jan 11&$L$ Spectroscopy&1200&0.66&HR 1931\\
      d216-0939&2006 Jan 15&$H$ Imaging&90&0.49&FS 13\\ 
      			&2006 Jan 15&$K$ Imaging&90&0.52&FS 13\\
      			&2006 Jan 15&$L^{\prime}$ Imaging&180&0.42&HD 40335\\
			&2006 Jan 15&$K$ Spectroscopy&480&0.44&HR 2328\\
      			&2006 Jan 15&$L$ Spectroscopy&2160&0.42&HR 2328\\
      d218-354&2001 Mar 11&$L^{\prime}$ Imaging&405&1.29&HD 36512\\
      			&2001 Mar 10&$K$ Spectroscopy&480&0.81&HR 1931\\
      			&2001 Jan 10&$L$ Spectroscopy&1400&0.90&HR 1931\\
       \cutinhead{Comparison Observation for Vicinity Stars}
	JW 370&2007 Sep. 06&$L^{\prime}$ Imaging&135&0.27&HD 225023\\
	           &2007 Sep. 06&$L$ Spectroscopy&1260&0.27&HR 1724\\
	\enddata
    \end{deluxetable}

Imaging data at $L^{\prime}$ were obtained for all the objects, and in addition imaging at $H$ and $K$ were carried out for d216-0939. For all the imaging, a box-shaped five-image dither pattern was used with a separation of 5\farcs0 and sky frames were obtained 300\arcsec west away from the objects. Spectroscopy for d121-1925, d182-332, d183-405, and d218-354 was conducted at the Cassegrain focus with a slit width of 0\farcs6, corresponding to a spectral resolution of $\lambda/\Delta\lambda$ $\sim$ 360 and 220 at $K$ and $L$, respectively. Only for d216-0939, the slit width of 0\farcs45 was used at the Nasmyth focus to obtain the spectral resolution of 430 and 260 at $K$ and $L$. The position angle of the slit was set to the perpendicular direction of the silhouette disk for d121-1925, d182-332, d183-405, and d218-354. In the case of d216-0939, the position angle of the slit was chosen to obtain the nearby star JW 671 \citep{jon88} simultaneously. Standard A-BB-A sequence was applied to subtract the sky emission with an offset width of 7\farcs0.

Regarding flat-fielding, imaging at $L^{\prime}$ used a sky flat. A flat field image at $H$ and $K$ and also flat field spectrum at $K$ and $L$ were taken with an integration sphere illuminated by a halogen lamp.

In addition two stars, JW 370 \citep{jon88} and JW 671, which are located at the vicinity around d121-1925 and d216-0939 respectively are observed for comparison. The observing parameters for JW 370 are included in Table~\ref{tbl-obslog}. The spectrum of JW 671 was obtained simultaneously with d216-0939 on the same slit.

Standard data reduction processing was applied using the IRAF software packages for flat fielding, sky subtraction, telluric correction, and wavelength calibration. The standard stars for telluric correction are listed in Table~\ref{tbl-obslog}. Applied photometric values at $L^{\prime}$ for HD 36512, HD40335, and HD 225023 are 5.54, 6.44, and 6.979 mag, respectively \citep{leg03}. The magnitudes at $H$ and $K$ for FS 13 are assumed to be 10.176 and 10.126 \citep{cas92}. For spectroscopic reference, the spectral type of HR 1931, HR 2328, and HR 1724 are O9.5V, A0Vn, and A0V, whose spectra are assumed to have a blackbody radiation with a temperature of 34,600K, 9,480K, and 9,480K, respectively. The wavelength calibration was performed using the telluric absorption lines in the spectrum. Since spectral region around hydrogen recombination lines (Br$\delta$; 1.94508\,$\micron$,  Br$\gamma$; 2.16611\,$\micron$, Pf$\epsilon$; 3.03919\,$\micron$, Pf$\delta$; 3.29698\,$\micron$, Pf$\gamma$; 3.74054\,$\micron$, Br$\alpha$; 4.05224\,$\micron$) shows insufficient sky subtraction due to high spatial variation of the strong nebular emission from the \ion{H}{2} region, those wavelengths are omitted from the spectra except for d216-0939 which has rather weak nebular emission. The hydrogen absorption of spectroscopic standard star HR 2328 is appropriately removed using the method developed by \citet{vac03}. The differences of the airmass between the target and the standard for all the spectroscopy were kept to be less than 0.04, while imaging data were taken with an airmass difference of $\leq$ 0.13. 

\section{RESULTS}

For imaging data, the aperture photometry was performed with a radius of 1\farcs5 except for the case of d218-354 at $L^{\prime}$ with a radius of 2\farcs5. The sky region for the determination of zero level was selected carefully for each target because the observed region is nebulous and confused by many sources. It is essential to have a good image size under $\leq$ 1\farcs0. Thanks to the good seeing conditions almost throughout the observing runs, the confusion from the nearby stars did not cause serious problems. The only exception is for the case of d218-354. The seeing condition was poor for the observation of this target, which caused worse quality of the data especially at $L^{\prime}$. Since the area of d218-354 is less confused by the nearby stars than the case of the others, there was no problem even in such bad seeing conditions. The photometric results are summarized in Table~\ref{tbl-phot} together with some of the values from the literature. Our infrared broad-band images do not show the silhouette disks, which are seen in the H$\alpha$ narrow-band images taken by $HST$. Therefore, the silhouette disks have a negligible effect on our photometric results of the continuum of the central sources. The $L^{\prime}$ images of d121-1925, d182-332, d183-405, and d218-354 show a point source profile, and the fluxes for those objects originate from the central source. Also in the case of d216-0939, which exhibits distinct scattered light in the $HST$ images, our infrared images at $H$, $K$, and $L^{\prime}$ do not clearly show the morphology of the circumstellar disk. The scattered light of the d216-0939 images is discussed in more detail in \S 4.

    \begin{deluxetable}{c c c c}
    	\tablecolumns{4}
	\tablewidth{0pc}
    	\tablecaption{Magnitudes of Targets\label{tbl-phot}}
	\tablehead{
     \colhead{Object}&\colhead{$H$}&\colhead{$K$}&\colhead{$L^{\prime}$}\\
     \colhead{}&\colhead{(mag)}&\colhead{(mag)}&\colhead{(mag)}}
      \startdata
      d121-1925&\nodata&11.74\tablenotemark{b}&10.83\\
      d182-316&13.10\tablenotemark{a}&12.25\tablenotemark{b}&11.18\\
      d183-405&12.80\tablenotemark{a}&12.24\tablenotemark{b}&10.93\\
      d218-354&9.69\tablenotemark{a}&9.49\tablenotemark{b}&9.00\\
      d216-0939&13.29&12.49&10.88\\
      \enddata
    \tablenotetext{a}{ \citet{mue02}.} 
    \tablenotetext{b}{ \citet{mcc96}.}
    \tablecomments{The typical 1 $\sigma$ uncertainties are 0.02, 0.02, and 0.1 mag for $H, K$, and $L^{\prime}$ in our data, respectively.}
    \end{deluxetable}

\begin{figure}
 \begin{center} 
 \includegraphics[scale=0.6]{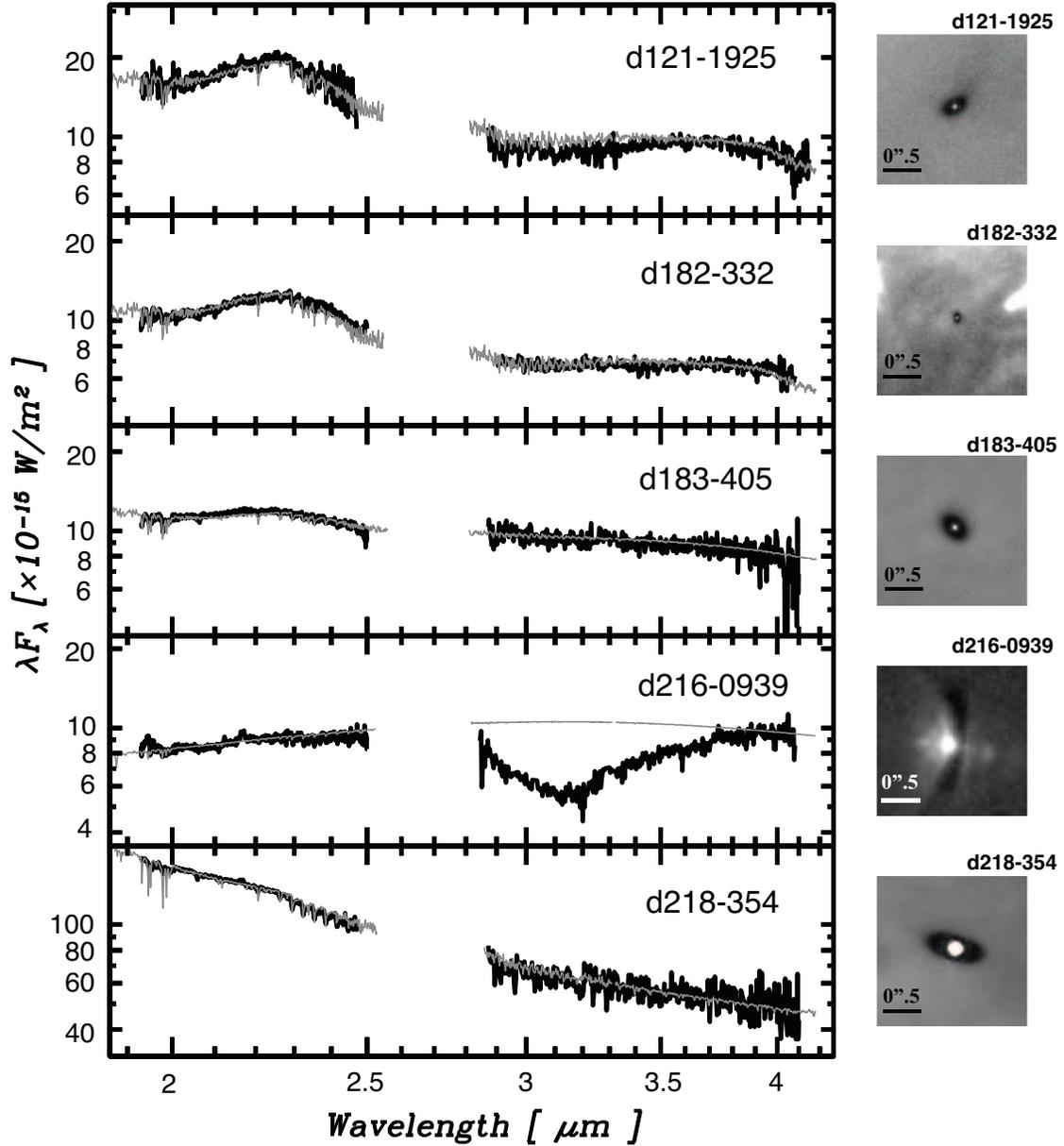}
  \caption{\label{fig-objspec}1.9--4.2\,$\micron$ spectra of five silhouette disks (black lines). The model spectra for fitting for the spectra of the targets are shown with gray lines. d216-0939 clearly demonstrates the broad absorption band around 3.1\,$\micron$, which is attributed to the water ice. The same feature can be marginally seen in the spectrum of d121-1925. The right panel shows the {\it HST} images taken with a narrow band filter of F656N at H$\alpha$ line. Images d121-1925, d182-332, d183-405, and d218-354 are from \citet{bal00}, and the d216-0939 image is from \citet{smi05}. All the figures are displayed in the same scale of 2\farcs73 $\times$ 2\farcs73, and the scale bar of 0\farcs5 is shown inside each image.
}
 \end{center}
\end{figure}

All the extracted 1.9--4.2\,$\micron$ spectra for the targets associated with the silhouette disk are shown with black lines of Figure~\ref{fig-objspec}. These spectra in the atmospheric windows of K-band (1.9--2.5\,$\micron$) and L-band (2.8--4.2\,$\micron$) were adequately offset to match to the photometric data determined in the above. Accurate photometric values at $K$ and $L^{\prime}$ are very critical to precisely extract the possible absorption feature around 3.1\,$\micron$ region. For d121-1925, d182-332, d183-405, and d218-354, the photometry at $K$ taken at the different epoch is used, and the possible variability of each target should be carefully taken into account. Regarding $K$ magnitudes for d182-332, d183-405, and d218-354, \citet{mcc96} and \citet{mue02} shows consistent values, which suggests those are not variable at $K$. As for d121-1925, there is no photometric data available at $K$ other than the one by \citet{mcc96} to investigate the possible variability. However, our spectroscopic continuum level shows sufficient continuity and it should be acceptable to assume that the $K$ magnitude measured by \citet{mcc96} is useful for the following estimate of the absorption feature. The derived continuum is confirmed to be in good agreement with the available photometry at $H$ except for d216-0939, which is brighter by 0.3 mag most probably due to contribution of the scattered light around the central source.

In Figure~\ref{fig-objspec}, the broad absorption feature around 3.1\,$\micron$ is definitely seen in the spectrum of d216-0939. For the other sources, the detailed determination of their continuum emission is obviously required. To determine the continuum level, the spectral template in the IRTF Spectral Library \citep{ray09} which is available on-line
\footnote{$\url{http://irtfweb.ifa.hawaii.edu/{\sim}spex/IRTF\_Spectral\_Library/}$}
is adopted. For d121-1925, d183-405, and d218-354, the spectral type and the visual extinction ($A_{V}$) which are summarized by \citet{get05} from the original data \citep{hil97,lum00} are applied as an initial assumption. The spectral type and the $A_{V}$ for d216-0939 are estimated by \citet{hil97}. All this information is summarized in Table~\ref{tbl-spt-av}. Meanwhile, no reliable spectral type is found for d182-332, and only $A_{V}$ by \citet{mcc96} is presented in the table for this object. 
To obtain the good continuum fitting, it is well known that some contribution of the infrared excess/veiling effect should be taken into account. Here, the single temperature blackbody radiation is added to reproduce the measured spectral continuum. This value represents the characteristic temperature of the material attributed to the longer wavelength continuum. In fitting the continuum, the parameters for the already known spectral type and $A_{V}$ were tried first. For d121-1925, d218-354, and JW 671, the spectra can not be fit using the already known parameters, and those were adjusted to obtain a good fit. The best fit parameters for the spectral model are shown in Table~\ref{tbl-spt-av}, and the derived optical depths of the water ice are presented in Table~\ref{tbl-tau}. Regarding the spectral type and $A_{V}$, the parameters that differ from previous works are represented with the bold font in Table~\ref{tbl-spt-av}. The detailed spectral classification is not covered by this paper, especially because the classification in the infrared is usually less accurate than in the optical. Since the discrepancy is small enough, it will not affect the following overall discussion.

\begin{deluxetable}{c c c c c c c c}
	\tabletypesize{\footnotesize}
	\tablecolumns{8}
	\tablewidth{0pc}
	\tablecaption{Spectral Type, Visual Extinction, Blackbody Temperature and Infrared Excess\label{tbl-spt-av}}
      \tablehead{
      \colhead{}         &\multicolumn{2}{c}{Previous Works}&\colhead{}&\multicolumn{4}{c}{Derived Parameter}\\
      \cline{2-3} \cline{5-8}\\
      \colhead{}&\colhead{Spectral}&\colhead{}&\colhead{}&\colhead{Spectral}&\colhead{}&\colhead{Blackbody}&\colhead{}\\
      \colhead{Object}&\colhead{Type}&\colhead{$A_{V}$}&\colhead{}&\colhead{Type}&\colhead{$A_{V}$}&\colhead{Temperature}&\colhead{$r_{L^{\prime}}$}\\
      \colhead{}&\colhead{}&\colhead{(mag)}&\colhead{}&\colhead{}&\colhead{(mag)}&\colhead{(K)}&\colhead{}}
      \startdata
      d121-1925&M4.5\tablenotemark{a}&5.83\tablenotemark{a}&&{\bf M6}&{\bf 6.8}&\nodata&0\\
       d182-332&\nodata&6.0--6.5\tablenotemark{c}&&M6&6.5&1000&0.13\\
       d183-405&M3\tablenotemark{b}&2.55\tablenotemark{b}&&M3&2.55&1100&0.29\\
       d216-0939&K5\tablenotemark{b}&0.72\tablenotemark{b}&&K5&0.72&1100&1.7\\
       d218-354&G6--K3\tablenotemark{b}&1.51\tablenotemark{b}&&{\bf M1}&{\bf 1.3}&500&0.29\\
      \cutinhead{Vicinity Stars}
	JW 370&K0--K3\tablenotemark{b}&5.36\tablenotemark{b}&&K1&5.36&1000&0.19\\
	JW 671&M2\tablenotemark{b}&1.72\tablenotemark{b}&&{\bf M6}&1.72&1500&0.14\\
      \enddata
 \tablenotetext{a}{ \citet{get05}.} 
 \tablenotetext{b}{ \citet{hil97}.}
 \tablenotetext{c}{ \citet{mcc96}.}
 \tablecomments{$r_{L^{\prime}}$ is defined as the ratio of the infrared excess at 3.77\,$\micron$ to the intrinsic stellar continuum flux: $F_{excess}$(3.77\,$\micron$)/$F_{intrinric}$(3.77\,$\micron$).}
\end{deluxetable}

\begin{deluxetable}{c c c c}
	\tablecolumns{4}
	\tablewidth{0pc}
    \tablecaption{Optical Depth of Water Ice Absorption at 3.1\,$\micron$\label{tbl-tau}}
	\tablehead{
      \colhead{Object}&\colhead{Peak}&\colhead{$\tau_{ice}$}&\colhead{$\Delta\nu$}\\        	\colhead{}&\colhead{($\micron$)}&\colhead{}&\colhead{(cm$^{-1}$)}
	}
      \startdata
      d121-1925&3.10	&0.12$\pm$0.01&443\\
      d182-316&\nodata&$\le$0.012&\nodata\\
      d183-405&\nodata&$\le$0.022&\nodata\\
      d218-354&\nodata&$\le$0.054&\nodata\\
      d216-0939&3.12&0.67$\pm$0.01&538\\
      \enddata
\end{deluxetable}

\begin{figure}
 \begin{center} 
  \includegraphics[scale=0.6]{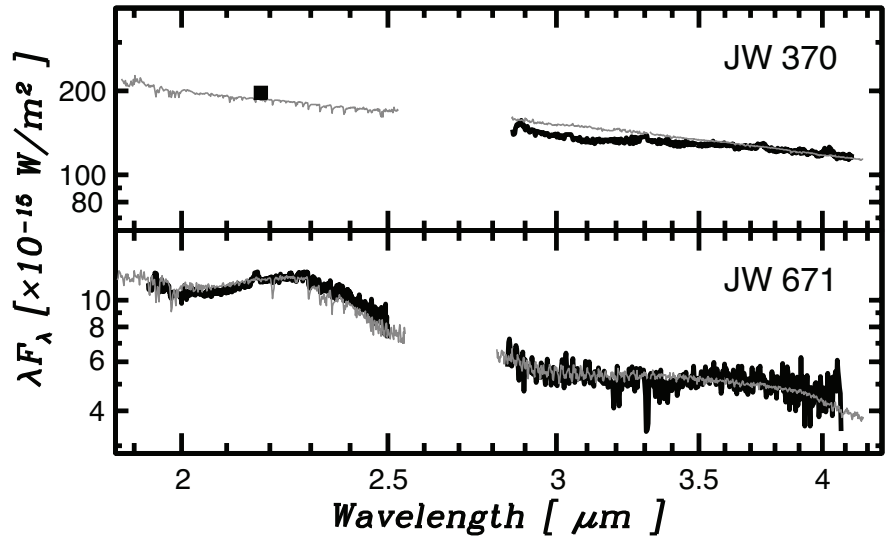}
  \caption{\label{fig-refspec}Spectra of JW 370 and JW 671. These are shown in the same range of 1.9--4.2\,$\micron$ for comparison. JW 370 and JW 671 are the vicinity stars of d121-1925 and d216-0939, respectively. The black filled square is the $K$ band flux measured from acquisition images.
}
 \end{center}
\end{figure}

For comparison in the discussion section below, the spectra of two stars: JW 370 and JW 671 are shown in Figure~\ref{fig-refspec} with an overplot of the estimated model spectra. Exactly the same procedure is applied for determination of the possible absorption around 3.1\,$\micron$ region. Since no spectral information in $K$ for JW 370 is obtained, the photometry data at $K$ is used for JW 370. The spectral type and the $A_{V}$ for those stars derived by \citet{hil97} can be found in Table~\ref{tbl-spt-av}.

\section{DISCUSSION}

\subsection{Origin of Detected Water Ice toward d121-1925 and d216-0939}
\begin{figure}
 \begin{center} 
  \includegraphics[scale=0.7]{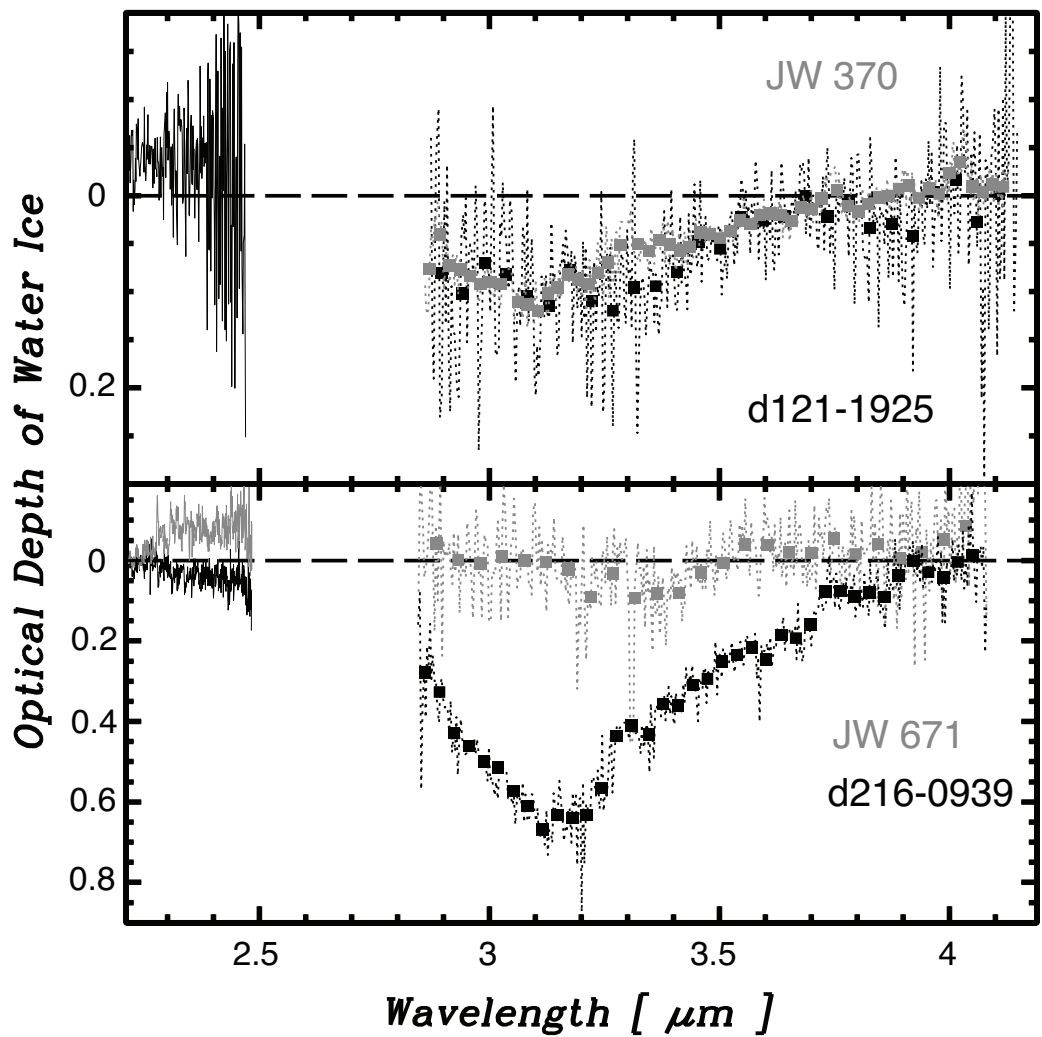}
  \caption{\label{fig-comp-spec}Comparison of the optical depths of the water ice toward the objects (d121-1925 and d216-0939) and the vicinity stars (JW 370 and JW 671, respectively). Since the depth and the profile are very similar between d121-1925 and JW 370, a common origin from foreground material is suggested for both water ice. On the other hand, the features are definitely different for the case of d216-0939 and JW 671. Localized water ice in the circumstellar disk of d216-0939 is responsible for the difference.
}
 \end{center}
\end{figure}

d121-1925 is located in one of the most embedded region ($A_{V} \geq 6$) in the Orion Nebula cluster \citep{sca11}, and contributions of the water ice material in the foreground to the detected water ice should be carefully evaluated. In fact, the shallow, but definite water ice absorption at 3.1\,$\micron$ was seen in the spectrum of JW 370 which is located only 3\farcs39 west (1403 AU in projected distance) away from d121-1925 in Figure~\ref{fig-refspec}. For comparison between those two detections of the water ice, Figure~\ref{fig-comp-spec} shows the optical depths of the detected water ice at 3.1\,$\micron$ for both d121-1925 and JW 370. The profile and the depth of those two features are found to be matched well within the signal-to-noise ratio, and it is most likely that the detected water ice in both d121-1925 and JW 370 could be attributed to the common origin of the foreground material.

In contrast, the \citet{sca11} data shows less extinction ($A_{V} \sim 3$) in the region where d216-0939 resides. The nearby star JW 671 is 4\farcs38 south-west (1910 AU in projected distance) away from d216-0939 and is used for evaluation of possible extension of the water ice absorbtion around the d216-0939 disk. Figure~\ref{fig-refspec} shows that JW 671 has no water ice absorption, and the difference of the optical depth of the water ice in d216-0939 and JW 671 is presented in Figure~\ref{fig-comp-spec}. JW 671 is known to be a member of the Orion Nebula cluster \citep{fur08}, and thus it is not a foreground star. Therefore, the water ice detected toward d216-0939 is concluded to be localized at least within a scale of 2000 AU from the central star, which is comparable to a typical size of the disk-envelope around young stars. 

\subsection{Possible Explanation for Water Ice Detection in d216-0939 Disk}

Silhouette disks in M42 and M43 represent a wide variety of characteristics in its mass, size, and inclination angle \citep{mcc96,smi05,bal00,ric08,man10}. In addition, the central stars associated with the silhouette disks also have range of spectral type, mass, and $A_{V}$ \citep{hil97,lum00} as well as various distance from the O, B stars (Trapezium cluster; mostly $\theta^{1}$ Ori C for M42 and NU Ori for M43) exciting the \ion{H}{2} regions \citep{vic05}.  Parameters of the disk mass, size, inclination angle, and distance from the exciting stars are summarized in Table~\ref{tbl-para}. In the following, the relationship of the water ice detection with each parameter is discussed.

\begin{deluxetable}{c c c c c c}
	\tabletypesize{\footnotesize}
	\tablecolumns{6}
	\tablewidth{0pt}
    \tablecaption{Disk Parameters\label{tbl-para}}
	\tablehead{
      \colhead{}&\colhead{}&\colhead{}&\colhead{Inclination}&\colhead{Exciting}&\colhead{}\\
      \colhead{Object}&\colhead{Mass\tablenotemark{a}}&\colhead{Diameter}&\colhead{Angle}&\colhead{Source}&\colhead{Distance}\\
       \colhead{}&\colhead{($\times$10$^{-2}$ M$_{\sun}$)}&\colhead{(\arcsec)}&\colhead{(\arcdeg)}&\colhead{}&\colhead{(\arcsec)}
	}
	\startdata
      d121-1925&0.74$\pm$0.07&0.8\tablenotemark{b}&51\tablenotemark{b}&$\theta^{1}$ Ori C&247.08\\
       d182-332&$<$0.37&0.3\tablenotemark{b}&55\tablenotemark{b}&$\theta^{1}$ Ori C&27.23\\
       d183-405&$<$0.20&0.7\tablenotemark{b}&39\tablenotemark{b}&$\theta^{1}$ Ori C&50.28\\
       d216-0939&4.53$\pm$0.06&2.6\tablenotemark{c}&75--80\tablenotemark{c}&NU Ori&400.49\\
       d218-354&2.37$\pm$0.04&1.4\tablenotemark{b}&65\tablenotemark{b}&$\theta^{1}$ Ori C&85.39\\
	\enddata
	\tablenotetext{a}{\citet{man10}.}
	\tablenotetext{b}{\citet{bal00}.}
	\tablenotetext{c}{\citet{smi05}.}
	\tablecomments{Distance is measured from the listed O, B stars responsible for excitation of the \ion{H}{2} region.}
\end{deluxetable}

\subsubsection{Disk Mass}
\citet{man09} derived the disk mass of (4.50 $\pm$ 0.06) $\times$ 10$^{-2}$ $M_{\sun}$ for d216-0939 from the submm measurement at 880\,$\micron$. This is the second most massive circumstellar disk among their large sample of protoplanetary disks in M42 and M43, and the most massive in our five targets.  In addition, the d216-0939 disk has the second largest size among all the silhouette disks in M42 and M43 \citep{vic05,smi05}. It might be straightforward to expect a larger column density of water ice in the bigger disk. However, taking into account the null detection of the water ice in the d218-354 disk which is the second most massive and the second largest in our sample, this is not a unique reason for finding water ice only in the d216-0939 disk. 

\subsubsection{Distance from O, B stars}
The distance to exciting sources for \ion{H}{2} regions is a key parameter for determining the water ice abundance, because strong UV radiation from the early type sources can dissociate water molecules located on the mantles of dust grains. This effect is known to arise as the threshold extinction ($A_{V;0}$) in relationship of  $N(H_{2}O)$ = $q$ $\times$ ($A_{V}$ - $A_{V;0}$) between column density of water ice ($N(H_{2}O)$) and visual extinction ($A_{V}$). Many studies have been done for determination of $A_{V;0}$ toward various molecular clouds such as Taurus, $\rho$ Ophiuchus, R CrA, and Serpens \citep{whi96,tei99} and $A_{V;0}$ is found to vary from 2 to 10 for those clouds due to the different internal UV radiation. \citet{tei99} found that the $q$ parameter is comparable between two molecular clouds in Taurus ((1.0 $\pm$ 0.1) $\times$ 10$^{17}$) and $\rho$ Ophiuhcus ((0.92 $\pm$ 0.04) $\times$ 10$^{17}$), and therefore we adopt $q$ = 1 $\times$ 10$^{17}$ hereafter to find the threshold extinction for our targets. 

JW 370 is located at a distance of 246\farcs4 from $\theta^{1}$ Ori C with $A_{V}$ = 5.36, and the water ice toward the object is detected with $\tau_{ice}$ = 0.12 and $\Delta\nu$ = 443\,cm$^{-1}$ in our data. The column density of the water ice is calculated as 2.66 $\times$ 10$^{17}$ cm$^{-2}$ using $N(H_{2}O)=\tau_{H_{2}O}\centerdot\Delta\nu/(2.0\times10^{-16})$ \citep{dhe86}. As result, the threshold extinction $A_{V;0}$ for JW 370 is estimated to be 2.70. 

Since d182-332, d183-405, and d218-354 are located much closer to the exciting O, B stars, much higher threshold extinction should be expected. In this sense, there is little chance for d183-405 and d218-354 to have the water ice mantle around the grains due to the lower $A_{V}$ in these sources (see Table~\ref{tbl-spt-av}). For the case of d182-332, higher extinction of $A_{V}$ = 6.0--6.5 was estimated by \citet{mcc96}. However, d182-332 is located very close to $\theta^{1}$ Ori C (27\farcs2) and the much higher threshold of $A_{V;0}$ could hamper the water ice formation onto the grains. 

d216-0939, which shows the moderate water ice absorption in its disk, has a significantly larger distance (400\farcs5) from NU Ori (exiting source of M43) compared to the other targets in our sample, and it should suffer least from intense UV radiation. Therefore, the threshold extinction of $A_{V;0}$ can be assumed to be no greater than 2.70. Whereas \citet{hil97} derived very small extinction ($A_{V}$ = 0.72), it is not sufficient to produce the water ice even with such low extinction threshold. The $A_{V}$ of 0.72 is likely to be underestimated due to contribution of the reflection nebula in the optical band. 

Lastly, d121-1925 has moderate distance from $\theta^{1}$ Ori C (247\farcs1) and the $A_{V;0}$ should be comparable to 2.70 around its vicinity star JW 370. The foreground water ice detected toward d121-1925 is consistent with the estimated $A_{V}$ of 5.83 \citep{hil97}. Taking into account the overall structure of M42 \citep{ode10}, the foreground materials should act as a good shield from the UV radiation to keep the water ice around the grains in the disk. Therefore, the null detection of water ice in the disk of d121-1925 needs to be explained in some other way.

\subsubsection{Inclination Angle of Disk}
The most plausible explanation for the water ice features in our targets is related to the disk geometry. \citet{pon05}, \citet{ter07}, and \citet{aik12} clearly detected the water ice features in the disk whose inclination is very high ($i$ $\geq$ 70$\arcdeg$) owing to the abundant materials for the absorption. \citet{pon05} pointed out that the disk inclination is an important parameter for the water ice detection in their disk model of CRBR 2422.8-3423. In their model, the inclination of 69 $\pm$ 3$\arcdeg$ seems to act as the critical angle. 

\begin{figure}
\begin{center} 
\includegraphics[scale=0.7]{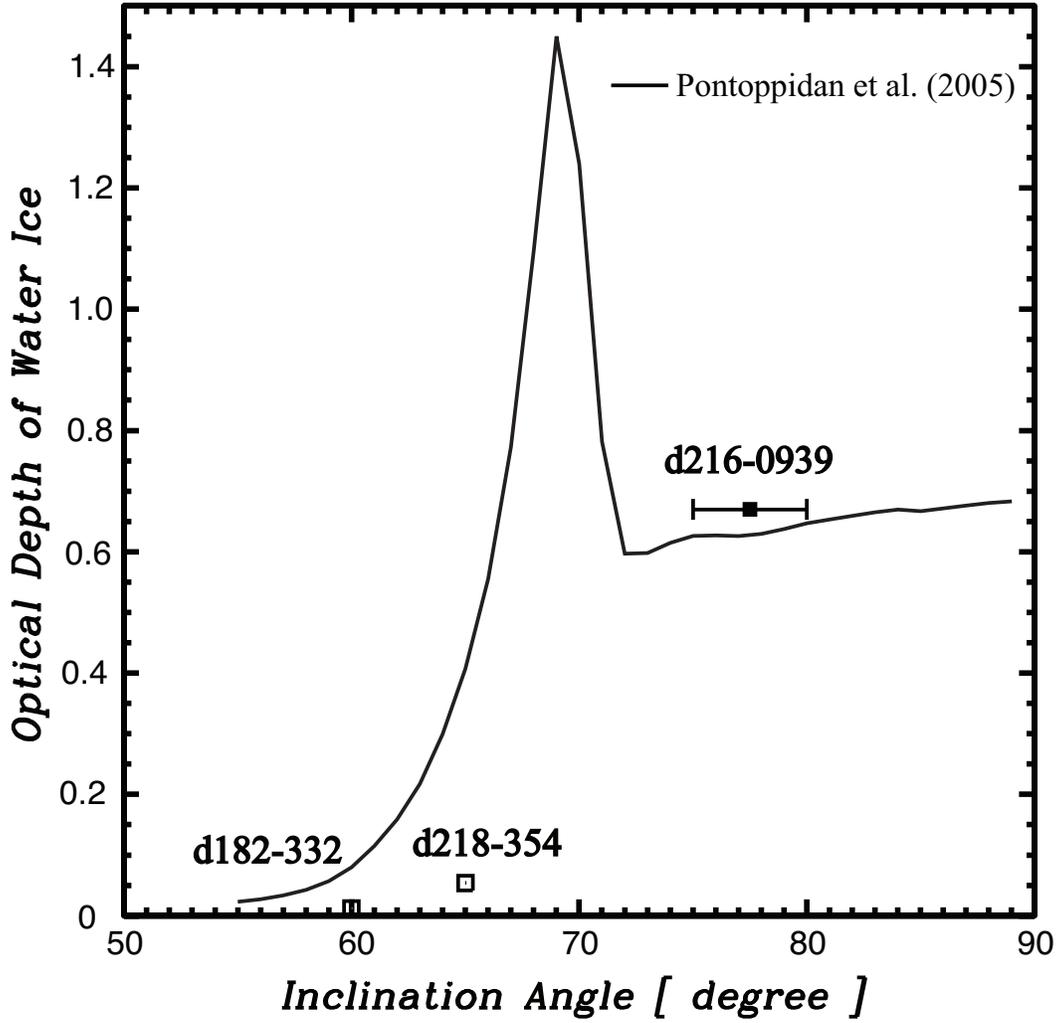}
 \caption{\label{fig-inctau}Optical depth of the water ice vs. inclination angle of the disk. Filled square with error bar shows the detection of the water ice for the d216-0939 disk. Open squares represent the estimated upper limit (1$\sigma$) for non-detection of the water ice for the d182-332 and d218-354. d121-1925 and d183-405 with lower inclination angle are not shown in this figure. Solid line is the expected optical depth of the water ice in the model for CRBR 2422-3423 disk derived by \citet{pon05}.
}
\end{center}
\end{figure}

In the disk with an inclination angle smaller than the critical angle, the expected optical depth is rapidly decreasing to zero. The two sources: d121-1925 and d183-405 are thought to have the small inclination angles of 51$\arcdeg$ and 39$\arcdeg$, respectively. No water ice detection in the disk of those objects is consistent with the Pontoppidan et al. model. For the remaining three sources: d182-332, d218-354, and d216-0939 with higher inclination angle of the disks, the observed optical depths of the water ice absorption are presented in Figure~\ref{fig-inctau} with an overplot of the $\tau_{3.08}$ shown by \citet{pon05}. In the case of d216-0939 disk with the inclination angle of 75--80$\arcdeg$ above the critical angle, the detected optical depth is explained well by this model. 

On the other hand, the question should arise for d182-332 and d218-354 in this scheme. As for d182-332, the expected optical depth is rather low ($<$ 0.1), and so systematic errors for spectral fitting and/or photometric measurements errors could be the cause of this discrepancy. The inconsistency for the case of d218-354 which has rather high inclination angle (65$\arcdeg$) seems more serious. This should be explained by a higher critical angle of the disk inclination and/or lower abundance of the water ice.  Considering that circumstellar disks so far detected exhibit a wide variety of morphology in scattered light, it is possible that the same critical angle of the disk inclination as in CRBR 2422.8-3423 should not necessarily be applied to all the disks. In fact, the d216-0939 disk seems to have a different critical angle of $\sim$ 77$\arcdeg$ judging from its flared disk, because the critical angle is expected to be determined by the opening angle of the disk. The problem with both the cases of d182-332 and d218-354 might be resolved with the optical depth curve shifted accordingly, if they have a higher critical angle of the disk inclination. Regarding the possible difference of the water ice abundance, we note that it should be affected significantly by much more intense UV field for the disks (d182-332, d183-405, and d218-354) located at the vicinity of O, B stars.

\subsection{Properties of d216-0939 Disk}

\subsubsection{Geometry for Light Path}

As mentioned in the above sections, \citet{smi05} revealed that the d216-0939 disk has an inclination angle of 75-80$\arcdeg$ which is well matched to the flared angle of the disk (77$\arcdeg$). This means that this target might provide with very unique opportunity for us to directly measure the absorption feature through the disk toward the central star. In the following, this suggested interesting geometry is re-examined using our near infrared data.

\begin{figure}
\begin{center} 
\includegraphics[scale=0.7]{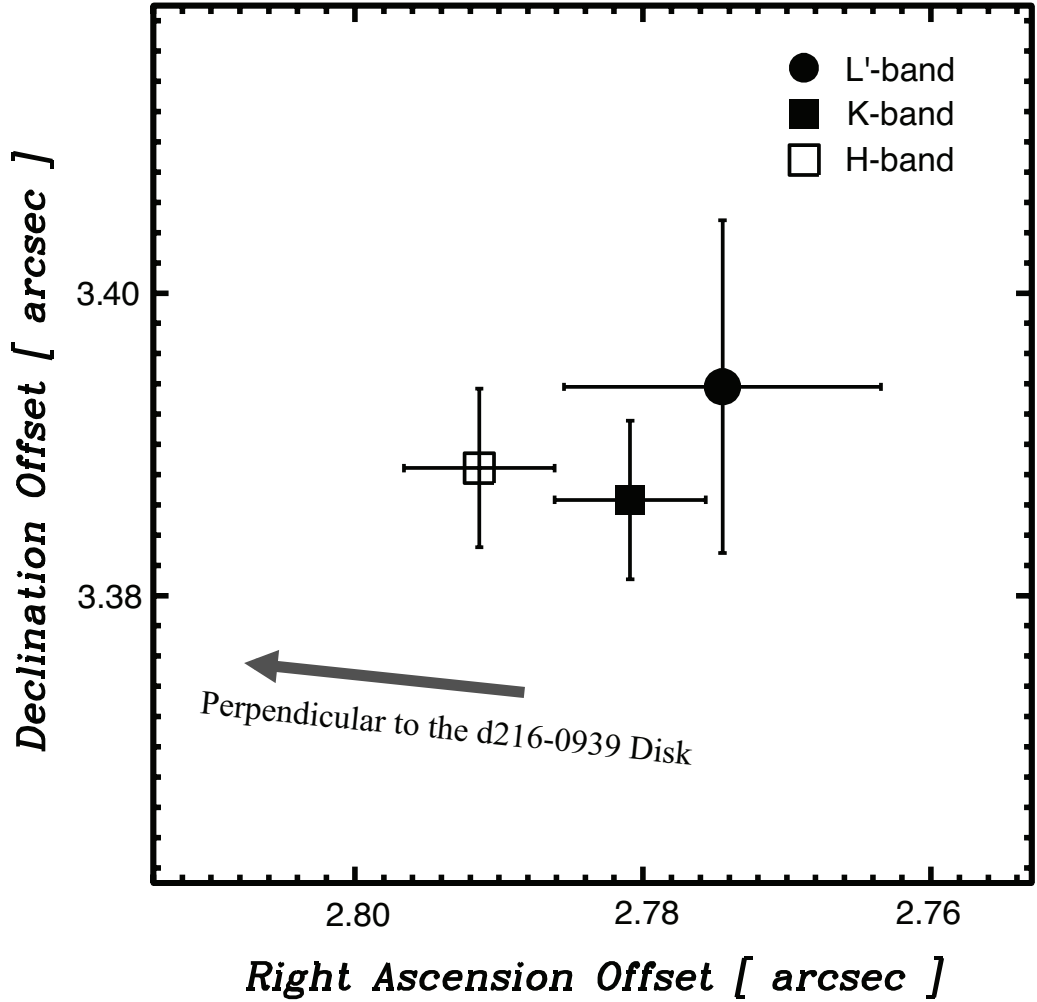}
 \caption{\label{fig-pos}Possible change of the peak position of the d216-0939 flux. The reference position is determined by the position of JW 671: the vicinity star of d216-0939. The peak at the longer wavelength shows the possible positional change toward the direction perpendicular to the disk. It may suggest that flux at shorter wavelength (at least $<$1.65\,$\micron$) comes from the scattered light by material around the star.
 }
\end{center}
\end{figure}

Using the optical depth parameters of $\tau_{ice}$ = 0.67 and $\Delta\nu$ = 538 cm$^{-1}$, the column density of the water ice toward d216-0939 is estimated to be 1.80 $\times$ 10$^{18}$ cm$^{-2}$ in the same way as the above. Since the detected water ice is known to reside in the disk region of d216-0939, the visual extinction caused by the disk material ($A_{V;disk}$) along a light-of-sight toward d216-0939 can be derived as 20.7 mag. This extinction is usually large enough to make the central source undetectable at the optical band. Thus, it is most likely that the previously measured optical flux could be considered to originate from the indirect reflected light from the central star. In fact, the position of the peak at $H$ and $K$ in our data shows a marginal offset into the direction perpendicular to the d216-0939 disk, while the peak position at $K$ and $L^{\prime}$ falls into the same area within the measurement accuracy (Figure~\ref{fig-pos}). This suggests that the flux shorter than 1.65\,$\micron$ appears as the reflected light. However, this result apparently conflicts with no detection of the positional shift of the peak flux at 0.550\,$\micron$ and 0.658\,$\micron$ bands reported by \citet{smi05}. Although it might be explained by the appropriate size of the large grains, this result is required to be addressed by using more sensitive and higher spatial resolution imaging in future.

\begin{figure}
\begin{center} 
\includegraphics[scale=0.7]{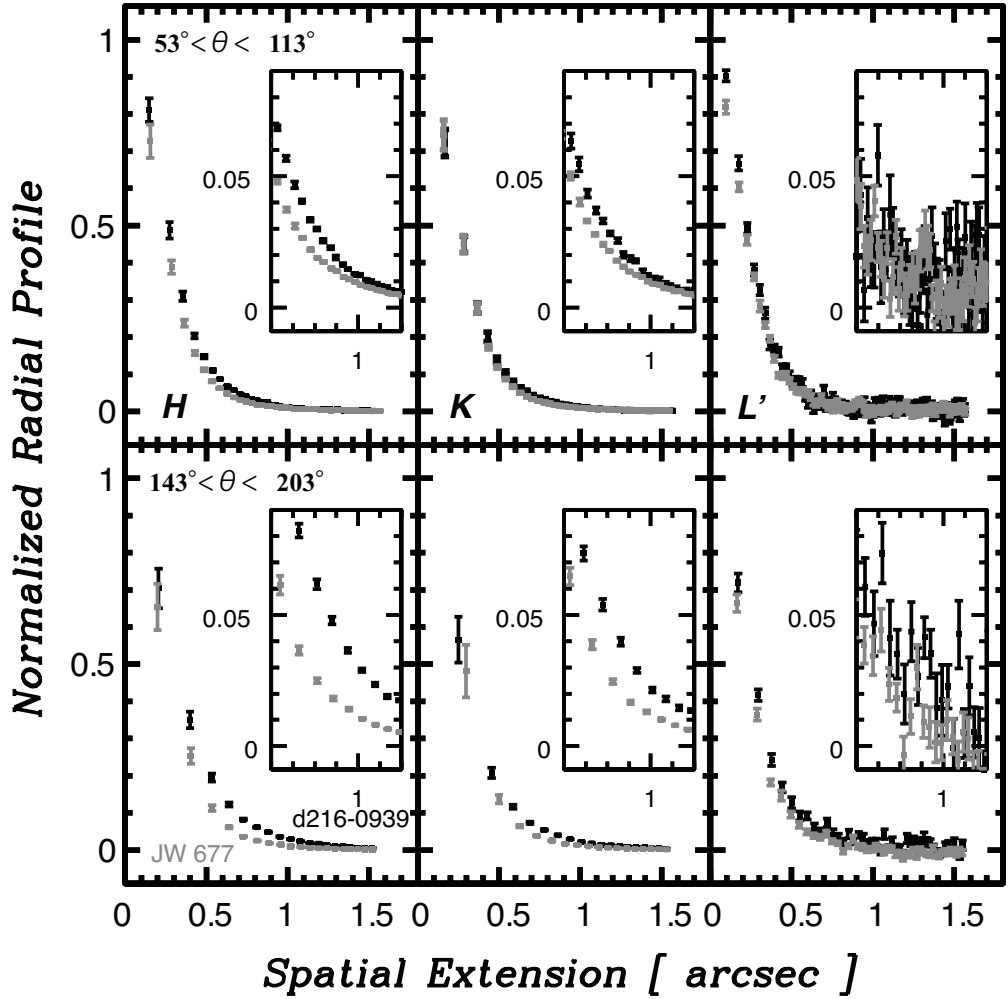}
 \caption{\label{fig-rad}Radial profiles of d216-0939 and JW 677. Upper panels are measured at the position angle of 53\arcdeg to 113\arcdeg, which is in the direction perpendicular to the d216-0939 disk. Lower panels are parallel to the disk direction with the position angle of 143\arcdeg to 203\arcdeg.
}
\end{center}
\end{figure}

Apart from the position offset, the radial profiles of d216-0939 at each band are investigated in Figure~\ref{fig-rad} to estimate the contribution of the reflected light to the flux.  This figure presents the radial profiles for both directions perpendicular to and along the d216-0939 disk (position angle: 83$\arcdeg$ and 173$\arcdeg$, respectively). JW 677 \citep{jon88} which is a brighter star located 10\farcs84 south is chosen as a  reference of the point spread function (PSF). In the profiles at $H$ and $K$, the extended features are clearly seen for both directions and the profiles along the disk appear more extended. Thus, the photometric results at $H$ and $K$ for d216-0939 include the scattered light component. On the other hand, at $L^{\prime}$, no significant extended feature can be observed for the direction perpendicular to the disk. The radial profile at $L^{\prime}$ along to the disk implies the possible extended flux. Higher signal-to-noise imaging with higher spatial resolution is required for concrete conclusion for this extended feature.

\subsubsection{Mass \& Density}

Considering the lack of change of the $K$ and $L^{\prime}$ peak position (Figure~\ref{fig-pos}) and the lack of evident difference in the $L^{\prime}$ profile of d216-0939 and JW 677 (Figure~\ref{fig-rad}), we assume that the $L^{\prime}$ light reaches us directly from the central star. The derived visual extinction, $A_{V;disk}$, and column density of the water ice, $N(H_{2}O)$, put direct constraints on physical parameters of the disk such as mass and density. Firstly, based on the relationship of $N_{H}$ = (1.59 $\times$ 10$^{21}$) $\times$ $A_{V}$ cm$^{-2}$ between $N_{H}$ and $A_{V}$ \citep{ima01}, $N_{H}$ is derived as 3.29 $\times$ 10$^{22}$ cm$^{-2}$. This number is in good agreement with the relationship between $N_{H}$ and the inclination angle of the silhouette disks in M42 found by \citet{kas05}. 

The silhouette disk of d216-0939 exhibits the clear structure of the disk, whose outer radius and scale height at the outer edge are measured to be 1\farcs3 (572 AU) and 0\farcs6 (264 AU), respectively \citep{smi05}. Thus, the path length can be estimated to be 8.56 $\times$ 10$^{15}$ cm and the averaged density can be calculated to be 6.38 $\times$ 10$^{-18}$ g cm$^{-3}$ accordingly. If this density is uniformly adopted to the entire disk, the total disk mass is calculated to be 1.94 $\times$ 10$^{-3}$ M$_{\sun}$. The derived mass is smaller by a factor of 20 than the one from the submm measurement. The difference can be naturally explained by the fact that the mid-plane of the circumstellar disk has higher density than the surface of the disk \citep[see for example][]{hog11}. 

\subsection{Nature of Water Ice in d216-0939 Disk}

\begin{figure}
\begin{center} 
\includegraphics[scale=0.7]{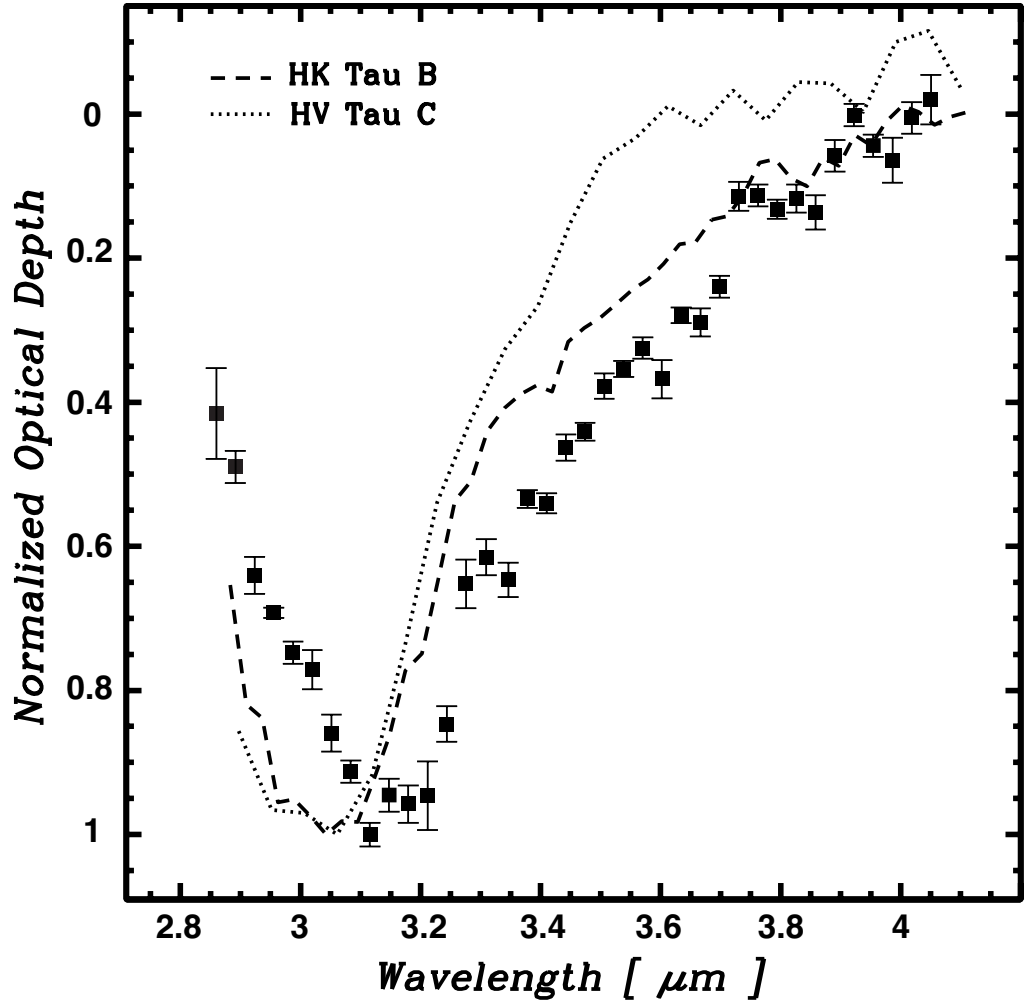}
 \caption{\label{fig-taucomp-eo}Comparison of the normalized optical depths of the water ice between d216-0939 and edge-on disk sources: HK Tau B and HV Tau C. Overall, the optical depth of the water ice in the d216-0939 disk is displaced to longer wavelengths.
}
\end{center}
\end{figure}

The most noticeable feature represented in the water ice absorption of the d216-0939 disk is the peak of the optical depth and its profile. In comparison to the water ice band of edge-on disks in Taurus molecular cloud shown in \citet{ter07}, the overall profile is shifted to longward wavelengths together with the peak absorption (Figure~\ref{fig-taucomp-eo}). In fact, the enhancement of the water ice feature around 3.2\,$\micron$ has never been seen in any astronomical sources before, and the cause of this feature could be very peculiar. The origin of this broad absorption in the red part ($\geq$ 3.4\,$\micron$) could be a scattering effect and/or other molecules such as solid methanol and ammonia hydrate \citep{dar02}. However, the determination of the red continuum adopted by a black body with a single temperature has some uncertainty and it is required to note the possibility that the optical depth at longer wavelengths may be overestimated.

Regarding the peak of the optical depth of the water ice band, it is well known that the higher degree of crystallization and larger grain size for water ice shift the optical depth peak to a longer wavelength \citep{smi89}. The fully crystallized water ice with a grain radii of $\leq$ 0.5\,$\micron$ shows a sharper peak around 3.1\,$\micron$ \citep[ex.][]{dar01}. There are several sources showing their peak around 3.1\,$\micron$, which is attributed to different degrees of crystallization. For the following discussion, it is noted that those sources do not show the 3.2\,$\micron$ absorption. For the case of the d216-0939, the absorption seen around 3.2\,$\micron$ is significantly enhanced, whereas the first peak position of the detected water ice is located around 3.1\,$\micron$ which agrees with the signature of the crystallized water ice. The peak shift of the optical depth for the fully crystallized water ice was presented by \cite{leg79}. In their work, the fully crystallized water ice with a grain size of 1\,$\micron$ shows the peak around 3.2\,$\micron$. 

\begin{figure}
\begin{center} 
\includegraphics[scale=0.7]{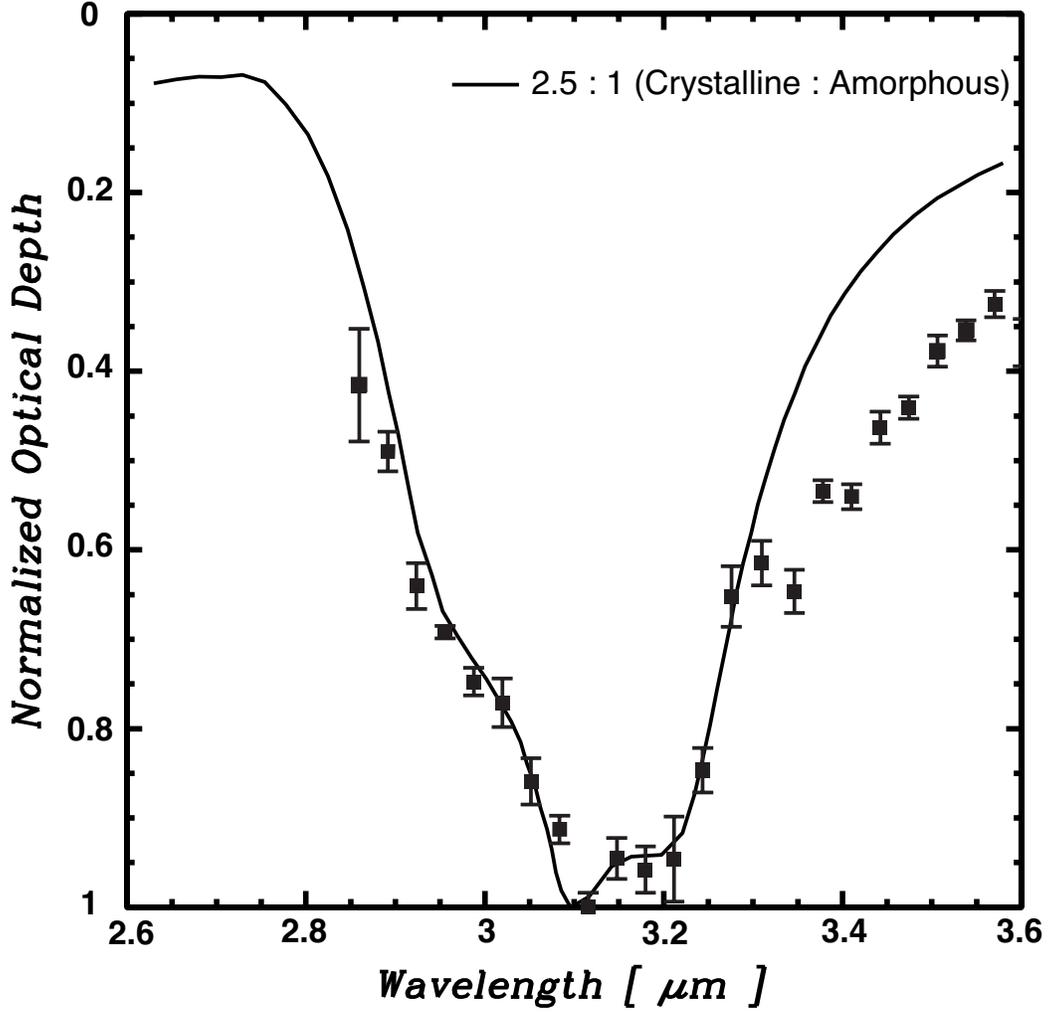}
 \caption{\label{fig-taucomp-model}Comparison of the normalized optical depth of the water ice between the model and the detection. Solid line shows composition of purely crystallized and pure amorphous water ice by 2.5:1. This model is found to match sufficiently to explain the absorption profile at wavelengths shorter than 3.3\,$\micron$. The extra absorption at 3.4\,$\micron$ seen in the spectrum could be attributed to a scattering effect and/or various possible carriers of the C-H bond such as CH$_3$OH ice.
}
\end{center}
\end{figure}

Taking into account fully crystallized water ice with such a large size ($\sim$1\,$\micron$), the model fitting to the profile of the optical depth of the water ice is considered with the following procedure, which is the same method with which \citet{sch10} have obtained a sufficient fit to the water ice features found in their sources. The optical depth of the water ice absorption is assumed to be a linear composition of the mass extinctions of amorphous, and partially and fully crystallized water ice with a grain size of 0.1, 0.5, 0.8, 1.5\,$\micron$. We used the grain model described by \citet{sch10} to calculate the water ice optical depth profile, and the mass extinction curves are derived using the program MIEX \citep{wol04} with an assumption of spherical and compact ice particles. Using the mass extinction parameters for each type of water ice, the best fit to the optical depth profile of the water ice detected in the d216-0939 disk is found to result from the combination of the large-sized (0.8\,$\micron$) fully crystallized water ice and the small-sized (0.1\,$\micron$) pure amorphous water ice with a mass ratio of 2.5:1. The best fit curve is presented in Figure~\ref{fig-taucomp-model}. The grain growth indicated by this result is predicted to occur in the protoplanetary disk. Therefore, this grain growth signature of the water ice feature strongly supports the origin of the detected water ice to be in the disk. As for the grain size of 0.8\,$\micron$, it is significantly smaller than the grain size (1.9--4\,$\micron$) found by \citet{shu03} for the largest silhouette disk d114-426. It may be due to different environments caused by UV radiation.

The detection of the fully crystallized water ice indicates that the detected water ice has experienced thermal processing to a temperature of $\sim$ 160K. Considering water ice particles are seen with a viewing angle of 75--80$\arcdeg$ \citep{smi05}, it should be located almost on the surface of the flared disk and the temperature of the water ice particles could be heated by radiation from the central star. \citet{sch10} interpreted that their spatially resolved detection of the crystallized water ice toward YLW 16A in $\rho$ Ophuichus molecular cloud comes from the outer surface of the disk. In our data, the water ice distribution in radial direction cannot be probed directly. Since the grain growth is expected to occur most effectively in the mid-plane of the circumstellar disk, it would be natural to consider some mixing process to exchange the materials between the mid-plane and the surface of the disk. The
mid-plane of the circumstellar disk is cold enough to keep water ice in amorphous form, and the detected large-size fully crystallized water ice at the surface of the disk likely originates from the heating of large-size amorphous water ice transferred from the mid-plane of the disk. In this scenario, the detected small-size amorphous water ice particles should reside at the colder and less dense outer region of the disk that is further away from the central star than the crystallized water ice.

\section{CONCLUSION}

We obtained 1.9--4.2\,$\micron$ spectra for five targets associated with the distinct silhouette disk. For three objects: d182-332, d183-405, and d218-354 no water ice feature was detected primarily due to lower angle of the disk inclination than required for detection of the water ice signature. In addition those objects are too close to the O, B stars exciting the \ion{H}{2} region.  Whereas d121-1925 exhibits a slight water ice band at 3.1\,$\micron$ with an optical depth of 0.12, a nearby star shows almost the same absorption as this object. Thus, the origin of those water ice absorption could be common, and it is most likely attributed to the foreground materials of the water ice. In contrast, the comparison star near d216-0939 shows no water ice feature, while the moderate absorption ($\tau_{ice}$ $\sim$ 0.67) of the water ice at 3.1\,$\micron$ toward the d216-0939 is definitely seen. Therefore, we concluded that the water ice toward d216-0939 should be localized in $\sim$2000 AU scale and that the material responsible for the absorption could be inside the disk of d216-0939. Due to the clear morphology and suggested ideal geometry for our line-of-sight, we directly derived the average density on the disk surface to be 6.38 $\times$10$^{-18}$ g cm$^{-3}$. Remarkably, the profile of the optical depth of the water ice shows the signatures of fully crystallized water ice with a grain size of $\sim$ 0.8\,$\micron$. Secondary composition of small amorphous, that is,  non-evolved water ice indicates that this material could exist in outer region of the circumstellar disk.

The water ice feature at 3.1\,$\micron$ is at a wavelength where diffraction-limited images can be easily obtained by using adaptive optics (AO) with the 8-m Subaru telescope. The spatial resolution of $\sim$ 0$\farcs$08 at 3.1\,$\micron$ is comparable to the resolution achieved by $HST$ in the optical. Future AO observations of the silhouette disks can help to reveal the morphology of the water ice absorption in more detail.

\acknowledgements
We thank all the Subaru telescope staff for maintaining the IRCS and supporting the observation. We also acknowledge the development efforts of the instrument to the development team members to realize these high sensitivity observations.

{}

\end{document}